\documentclass{article}

\usepackage{arxiv}

\usepackage[utf8]{inputenc} % allow utf-8 input
\usepackage[T1]{fontenc}    % use 8-bit T1 fonts
\usepackage{hyperref}       % hyperlinks
\usepackage{url}            % simple URL typesetting
\usepackage{booktabs}       % professional-quality tables
\usepackage{amsfonts}       % blackboard math symbols
\usepackage{nicefrac}       % compact symbols for 1/2, etc.
\usepackage{microtype}      % microtypography
\usepackage{lipsum}		% Can be removed after putting your text content
\usepackage[authoryear,square,comma]{natbib}
\usepackage{doi}

\usepackage[T1]{fontenc}
\usepackage[pdftex]{graphicx} 
\usepackage{siunitx}

\usepackage{amsmath,mathtools}
\usepackage{amsfonts}
\usepackage{algorithm}
\usepackage[noend]{algpseudocode}
\usepackage{eqparbox}
\usepackage{accents}
\newcommand{\ubar}[1]{\underaccent{\bar}{#1}}

\newdimen{\algindent}
\algnewcommand\LeftComment[2]{%
\hspace{#1\algindent}$\triangleright$ \eqparbox{COMMENT}{#2} \hfill %
}

\usepackage{graphicx,amsmath}
\newcommand{\colvec}[2][.8]{%
  \scalebox{#1}{%
    \renewcommand{\arraystretch}{.8}%
    $\begin{bmatrix}#2\end{bmatrix}$%
  }
}

\newcommand{\vect}[1]{\mathbf{#1}}
\DeclareMathSymbol{\shortminus}{\mathbin}{AMSa}{"39}

\DeclareMathOperator{\rank}{rank}

\title{ Empirical Individual State Observability}

\author{ \href{https://orcid.org/0000-0002-0609-7662}{\includegraphics[scale=0.06]{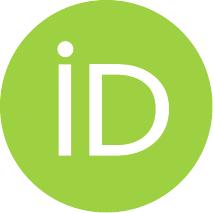}\hspace{1mm}Benjamin ~Cellini}\\
	Department of Mechanical Engineering \\
	University of Nevada, Reno \\
	Reno, NV 89557 \\
	\texttt{bcellini@unr.edu} \\
	\And
	\href{https://orcid.org/0000-0002-3268-1467}{\includegraphics[scale=0.06]{orcid.pdf}\hspace{1mm}Burak ~Boyacıoğlu} \\
	Department of Mechanical Engineering \\
	University of Nevada, Reno \\
	Reno, NV 89557 \\
	\texttt{bboyacioglu@unr.edu} \\
	\And
	\href{https://orcid.org/0000-0001-6538-7179}{\includegraphics[scale=0.06]{orcid.pdf}\hspace{1mm}Floris ~van Breugel} \\
	Department of Mechanical Engineering \\
	University of Nevada, Reno \\
	Reno, NV 89557 \\
	\texttt{fvanbreugel@unr.edu} \\
}

\renewcommand{\shorttitle}{Empirical Individual State Observability}

\hypersetup{
pdftitle={Empirical Individual State Observability (Preprint)},
pdfauthor={Benjamin~Cellini, Burak~Boyacioglu, Floris~van~Breugel},
}

\begin{document}
\maketitle

\begin{abstract}
A dynamical system is observable if there is a one-to-one mapping from the system's measured outputs and inputs to all of the system's states. Analytical and empirical tools exist for quantifying the (full state) observability of linear and nonlinear systems; however, empirical tools for evaluating the observability of individual state variables are lacking. Here, a new empirical approach termed Empirical Individual State Observability (E-ISO) is developed to quantify the level of observability of individual state variables. E-ISO first builds an empirical observability matrix via simulation, then determines the subset of its rows required to estimate each state variable individually. We present a convex optimization approach to do this efficiently. Finally, (un)observability measures for these subsets are calculated to provide independent estimates of the observability of each state variable. Multiple example applications of E-ISO on linear and nonlinear systems are shown to be consistent with analytical results. Broadly, E-ISO will be an invaluable tool both for designing active sensing control laws or optimizing sensor placement to increase the observability of individual state variables for engineered systems, and analyzing the trajectory decisions made by organisms.
\end{abstract}

% keywords can be removed
\keywords{ Observers for nonlinear systems \and Numerical algorithms  \and Sensor fusion}

\section{Introduction}
State estimation is a critical component of many tasks involving dynamic systems. A prerequisite for accurate estimation is that a system's states are observable, i.e. that there is a one-to-one mapping from a system's measured outputs and inputs to its states. Most estimation methods (e.g. a Kalman filter) rely on all of a system's states being observable and will otherwise fail in most cases \cite{Li2019}. Mathematical tools have been developed to assess the observability of systems and therefore help us understand how well an estimator will perform \cite{Zeng2018, Southall1998, mei2022mobile}. For nonlinear systems, where the observability may depend on the current states and control inputs, these tools can inform the design of control laws to achieve state trajectories and/or sensor locations or sensor configurations that improve the observability of the system.

However, certain control tasks only require estimating a single state variable (or a subset), instead of the full state vector. For instance, given a (nonlinear) system with unknown model parameters, it may be desirable to only estimate these parameters periodically, while ensuring continuous observability of other states. In some high-dimensional systems, such as fluid-structure interactions, it may only be necessary to estimate specific states for purposes of control \cite{hickner2023data}. Many navigation tasks also only require partial observability to achieve a desired outcome. A classic example is proportional navigation, a guidance law to ensure a collision course that only requires estimating one state: absolute bearing angle \cite{murtaugh1966fundamentals}.
% , yanushevsky2018modern, fabian2018interception, brighton2017terminal
%A classic example is proportional navigation, a guidance law \cite{murtaugh1966fundamentals} used by many homing missiles \cite{yanushevsky2018modern}, predatory insects \cite{fabian2018interception}, and birds of prey \cite{brighton2017terminal} to ensure a collision course with a moving target by controlling the thrust angle to maintain the target at a constant absolute bearing without requiring estimates of velocity. Similar rules are used by ship captains \cite{maloney1994chapman} and insects \cite{zabala2012simple} to avoid collisions. 
Flying insects engaged in chemical plume tracking behaviors \cite{van2014plume} may also prioritize estimating ambient wind direction over other states such as ground speed. Prior work has shown that the stereotyped zigzagging flight trajectories flying insects use may be tuned to enhance the observability of the high-priority state (ambient wind direction) \cite{VanBreugel2021a, VanBreugel2022}. In each of these examples, it is critical to have tools for evaluating the observability of individual state variables, instead of the full state vector.

Although analytical tools exist for determining if a particular state variable of a dynamic system is observable \cite{VanBreugel2021a,Anguelova2004}, or if a linear combination of states is functionally observable \cite{fernando2010functional}, there are no established methods to quantify the level of observability. Attempts to accomplish this task include using measures based on eigenvalues of the linear observability Gramian \cite{marques1986relative}, the singular values of the observability matrix \cite{yim2002autonomous, rui2008observable}, filter performances  \cite{zhigang2015adaptive}, and the estimation ambiguity \cite{gong2020partial}. These analytical methods become impractical when the dynamic model is missing or sometimes when the dynamics are nonlinear. As an alternative, empirical methods have been developed for quantifying a system's observability. The advantage of empirical tools, such as the empirical observability Gramian \cite{Lall1999}, is that they do not rely on an analytical model of the system and only require the ability to simulate it. However, how to tease out the relative observability of each state variable remains unclear. To our knowledge, an empirical method to quantify the observability of individual state variables does not exist.

This work presents a derivative-free empirical framework for evaluating the observability of individual states: Empirical Individual State Observability (E-ISO). Whereas prior approaches primarily employ the empirical observability Gramian, E-ISO utilizes the empirical observability \textit{matrix}. First, analytical observability tools are reviewed. Then, a framework for constructing an empirical observability matrix is presented. 
%From here, E-ISO applies convex optimization to efficiently collect a subset of rows in the observability matrix necessary for reconstructing each state individually. 
From here, E-ISO selects a subset of rows from the observability matrix that are necessary for reconstructing each state individually, and a convex optimization framework to efficiently perform this selection is presented.
Lastly, measures of (un)observability are calculated for each subset of rows, yielding an independent measure of (un)observability for each state variable. By considering example linear and nonlinear systems, it is shown that E-ISO can facilitate both sensor selection and trajectory planning to increase the observability of specific state variables.

\section{Nonlinear Observability Background}
To relate our E-ISO method to existing tools, we begin with a brief review of analytical and empirical observability. %analyses of discrete- and continuous-time nonlinear systems.
\subsection{Analytical Observability: Review}
Observability is a fundamental system property that characterizes the existence of an one-to-one (injective) mapping from measurements to state space with the knowledge of inputs. For nonlinear systems, we discuss weak observability, i.e. distinguishability of the unknown initial state in an open neighborhood based on finite-time measurements and input information \cite{nijmeijer1990nonlinear}.

Consider the continuous-time/discrete-time nonlinear time-invariant system dynamics,
 \begin{equation}\label{wo_delay}
 \small
  \Sigma_\text{c}:~
  \begin{aligned}
{\dot{\vect{x}}}(t) &=\vect{f}(\vect{x}(t),\vect{u}(t))\\
   \vect{y}(t) &= \vect{h}(\vect{x}(t)),
\end{aligned}
~ / ~  \Sigma_\text{d}:~
  %\begin{split}
   %{\vect{x}}_{k+1} &= A\vect{x}_k+B\vect{u}_k\\
   %\vect{y}_k &= C\vect{x}_k,\\
  %\end{split}
  \begin{aligned}
{\vect{x}}_{k+1} &= \vect{f}(\vect{x}_k,\vect{u}_k)\\
   \vect{y}_k &= \vect{h}(\vect{x}_k),
\end{aligned}
\end{equation}
where states $\vect{x}$ take values in a smooth $n$-dimensional state manifold $\mathbf{X}$, control inputs $\vect{u}$ take values in a subset $\mathbf{U}$ of an $m$-dimensional manifold $\mathcal{U}$, and outputs $\vect{y}$ take values in $\mathbb{R}^p$. $\vect{f_u}\coloneqq\vect{f}(\cdot, \mathbf{u})$ is a smooth vector field for each $\mathbf{u}\in \mathbf{U}$, and $\vect{h}=\begin{bmatrix}
       h_1&h_2&\cdots&h_p
   \end{bmatrix}^\top$ is the smooth output map of the system from $\mathbf{X}$ to $\mathbb{R}^p$. Given some control $\vect{u}^\ast$, the first $(w\shortminus1)$ time-derivatives of the output for the continuous-time system with $wp\geq n$ are given by
\begin{equation}\label{ctMapping}
\small
\begin{gathered}
    \begin{bmatrix}
        \vect{y}\\\vect{y}'\\\vect{y}''\\\vdots\\\vect{y}^{(w\shortminus1)}
    \end{bmatrix}=\begin{bmatrix}
    \vect{h}(\vect{x}(t))=L^0_\vect{f_{\vect{u}^\ast}}\vect{h}\\\vect{h}'(\vect{x}(t))=L^1_\vect{f_{\vect{u}^\ast}}\vect{h}\\\vect{h}''(\vect{x}(t))=L^2_\vect{f_{\vect{u}^\ast}}\vect{h}\\\vdots\\\vect{h}^{(w\shortminus1)}(\vect{x}(t))=L^{w\shortminus1}_\vect{f_{\vect{u}^\ast}}\vect{h}    
    \end{bmatrix}\coloneqq\mathcal{G}_c(w,\vect{x}(t),\vect{u}^\ast),
    \end{gathered}
\end{equation}
where the Lie derivative, $L_{\vect{f_{\vect{u}^\ast}}}\vect{h}$, denotes the derivative of $\vect{h}$ with respect to $\vect{x}$ on the vector field $\vect{f_{\vect{u}^\ast}}$, i.e.
$L_{\vect{f_{\vect{u}^\ast}}} \vect{h}=\frac{\partial \vect{h}}{\partial \vect{x}}\vect{f_{\vect{u}^\ast}}$,
% \begin{align}
%      L_{\vect{f_{\vect{u}^\ast}}} \vect{h}=\frac{\partial \vect{h}}{\partial \vect{x}}\vect{f_{\vect{u}^\ast}},
% \end{align}
and repeated Lie derivatives are calculated as $L_{\vect{f_{\vect{u}^\ast}}}^k \vect{h}=     L_{\vect{f_{\vect{u}^\ast}}}L_{\vect{f_{\vect{u}^\ast}}}^{k-1} \vect{h}$. The invertibility of the mapping $\mathcal{G}_c$ at a given state vector $\vect{x_0}\in\mathbb{R}^n$ requires its Jacobian to have the same rank as the dimension of the state space at $\vect{x_0}$, i.e. if the observability matrix ${\mathcal{O}}_{c,w}\coloneqq d\mathcal{G}_c=\frac{\partial \mathcal{G}_c(w,\vect{x},\vect{u}^\ast)}{\partial \vect{x}}|_{\vect{x}=\vect{x_0}}$ is full column rank, $\Sigma_\text{c}$ is observable \cite{sontag1984concept}.

Similarly, given an input sequence $\vect{u}^\dag=(\vect{u}_0,\vect{u}_{1},\dots,\vect{u}_{w\shortminus1})$,
$w$ consecutive measurements from the discrete-time system dynamics would give
\begin{equation}\label{dtMapping}
\small
\begin{aligned}
        \begin{bmatrix}
        \vect{y}_k\\\vect{y}_{k+1}\\\vect{y}_{k+2}\\\vdots\\\vect{y}_{k+w\shortminus1}
    \end{bmatrix}&=\begin{bmatrix}
        \vect{h}(\vect{x}_k)\\\vect{h}(\vect{x}_{k+1})=\vect{h}\circ \vect{f}_{\vect{u}_0}(\vect{x}_k)\\\vect{h}(\vect{x}_{k+2})=\vect{h}\circ \vect{f}_{\vect{u}_1}(\vect{x}_k)\circ \vect{f}_{\vect{u}_0}(\vect{x}_k)\\\vdots\\\vect{h}(\vect{x}_{k+w\shortminus1})=\vect{h}\circ \vect{f}_{\vect{u}_{w\shortminus1}}(\vect{x}_k)\circ\cdots \circ \vect{f}_{\vect{u}_0}(\vect{x}_k)
    \end{bmatrix}\\&:=\mathcal{G}_d(w,\vect{x}_k,\vect{u}^\dag),
\end{aligned}
\end{equation}
where $\circ$ denotes function composition. If the mapping $\mathcal{G}_d$ at $\vect{x_0}$ is invertible, then the discrete-time system is said to be $w$-step observable at $\vect{x_0}$, that is, the observability matrix ${\mathcal{O}}_{d,w}\coloneqq d\mathcal{G}_d=\frac{\partial \mathcal{G}_d(w,\vect{x},\vect{u}^\dag)}{\partial \vect{x}}|_{\vect{x}=\vect{x_0}}$ being full column rank implies $w$-step observability of $\Sigma_\text{d}$ \cite{moraal1995observer}.

One can also check the observability of a particular state variable by augmenting $\mathcal{O}_c$ or $\mathcal{O}_d$ with the basis vector corresponding to the state of interest $\vect{e}_j\in\mathbb{R}^n$ (e.g. $\vect{e}_1 = \begin{bmatrix} 1 & 0 & 0 \end{bmatrix}^\top$ for the first state of a three-state system) and checking if the rank changes. If $\rank(\begin{bmatrix}
    \mathcal{O}^\top&\vect{e}_j
\end{bmatrix}) = \rank(\mathcal{O}^\top)$, then the information required to obtain the state is already contained within $\mathcal{O}$, thus the $j^{th}$ state is observable. If the rank does change, then new information about the state was added to $\mathcal{O}$, thus the state is unobservable \cite{VanBreugel2021a}.

\subsection{Empirical Observability: Review}
Although analytical observability tools are valuable for systems with a known model, analytically obtaining the observability matrix is not always possible due to requirements like differentiability. For such systems, the empirical observability Gramian was introduced \cite{Lall1999}. Here, we show how to build an empirical observability matrix and relate it to the observability Gramian.

An empirical continuous-time observability matrix can be obtained by numerically computing ${\mathcal{O}} _{c,w}$. However, calculating the higher-order (time) derivatives that appear in the Jacobian of Eq. \ref{ctMapping} using difference formulas can be unreliable \cite{VanBreugel2020}. Hence, we focus on building an empirical observability matrix from discrete-time measurements.

Let $\vect{x_0}$ be the initial state vector of interest of the observability analysis, and let $\vect{u}^\dag$ be a nominal input. To construct a $w$-step empirical observability matrix, we perturb each initial state variable in positive and negative directions with a perturbation amount $\varepsilon$, that is, we simulate the given system dynamics $2n$ times in total and define the perturbed system's output vectors at time $k$ as:

\begin{equation}\small
\vect{y}^{\pm j}_k(\vect{x_0},\vect{u}^\dag,\varepsilon) =\vect{y}_k(\vect{x_0}\pm\varepsilon\vect{e_j},\vect{u}^\dag).
\end{equation}
Then the $w$-step empirical discrete-time observability matrix can be obtained as:
\begin{equation}\label{DT-EOM}\small
    {\mathcal{O}}_{d,w,\varepsilon}=\frac{1}{2\varepsilon} \begin{bmatrix}
        \Delta \vect{y}^1_0&\Delta \vect{y}^2_0&\cdots &\Delta \vect{y}^n_0\\
        \Delta \vect{y}^1_1&\Delta \vect{y}^2_1&\cdots &\Delta \vect{y}^n_1\\
        %\Delta \vect{y}^n_2&\cdots &\Delta \vect{y}^n_2\\
        \vdots &\vdots &\ddots & \vdots\\
        \Delta \vect{y}^1_{w\shortminus1}&\Delta \vect{y}^2_{w\shortminus1}&\cdots &\Delta \vect{y}^n_{w\shortminus1}
    \end{bmatrix},
\end{equation}
where $\Delta \vect{y}^j_k$'s are the differences between the output vectors at time $k$ for the perturbed state $j$, $\vect{y}^{+j}_k(\cdot)$ and $\vect{y}^{-j}_k(\cdot)$, i.e.
\begin{equation}\small
    \Delta \vect{y}^j_k(\vect{x_0},\vect{u}^\dag,\varepsilon)=\vect{y}^{+j}_k(\vect{x_0},\vect{u^\dag},\varepsilon)-\vect{y}^{-j}_k(\vect{x_0},\vect{u}^\dag,\varepsilon).
\end{equation}
For the purposes of evaluating observability along a state trajectory, it is convenient to construct the observability matrix in sliding windows assuming no noise (Fig. \ref{fig:O_method}A). 

Finally, the continuous-time observability Gramian  for the time interval $[0,w\Delta t]$ is defined as \cite{Georges-IDRIM-2020}:
\begin{equation}\label{Gramian1} \small
W_{\mathcal{O}_c}(0,w\Delta t)=%\int_0^T\prescript{1}{}{\mathcal{O}}_{c,\varepsilon}^{\top}\prescript{1}{}{\mathcal{O}}_{c,\varepsilon}dt\\
\int_{0}^{w\Delta t} \partial_{\vect{x_0}} \vect{y}^\top(\tau) \partial_{\vect{x_0}} \vect{y}(\tau) d\tau,
%\lim_{\substack{\Delta t\to 0 \\ w\Delta t=T}}
\end{equation}
and it can be shown that ${\mathcal{O}}_{d,w,\varepsilon}^{\top} {\mathcal{O}}_{d,w,\varepsilon}\Delta t$ with constant $w\Delta t$ would converge to $W_{\mathcal{O}_c}(0,w\Delta t)$ as the perturbation amount $\varepsilon$ and the discretization time step size $\Delta t$ go to zero, that is,
\begin{equation} \label{convOTO2G}\small
W_{\mathcal{O}_c}(0,w\Delta t)
\approx {\mathcal{O}}_{d,w,\varepsilon}^{\top} {\mathcal{O}}_{d,w,\varepsilon}\Delta t,
\end{equation}
for small $\varepsilon,\Delta t$. Hereafter, we simplify $\mathcal{O}_{d,\varepsilon}$ and $W_{\mathcal{O}_c}$ to $\mathcal{O}_\varepsilon$  and $W_{\mathcal{O}}$, respectively, and we use a value of $\varepsilon=10^{-3}$, considering that the standard size of each state variable is of order one.
%The $\frac{\Delta t}{4 \varepsilon}$ term is added to normalize the Gramian with respect to the time step of the simulation $\Delta t$.
\subsection{Unobservability Measures: Review}

%The observability and controllability Gramians are at least positive semidefinite by construction, that is, its eigenvalues are equal to its singular values. Gramian-based observability metrics were first suggested for linear systems \cite{muller1972}. 
%Two established metrics for quantifying the level of a nonlinear system's unobservability include the reciprocal of the minimum eigenvalue of $W_\mathcal{O}^c$, $1/\ubar{\lambda}(W_\mathcal{O}^c)$, and the condition number of the same matrix, $\kappa(W_\mathcal{O}^c)= \bar{\lambda}(W_\mathcal{O}^c)/\ubar{\lambda}(W_\mathcal{O}^c)$, and they are called unobservability index and estimation condition number, respectively \cite{Krener2009}. 
Since observability is determined by the invertibility of $W_{\mathcal{O}}$, established measures for quantifying the level of a nonlinear system's unobservability include the reciprocal of the minimum eigenvalue of $W_{\mathcal{O}}$, $1/\ubar{\lambda}(W_{\mathcal{O}})$, and the condition number of the same matrix, $\kappa(W_{\mathcal{O}})= \bar{\lambda}(W_{\mathcal{O}})/\ubar{\lambda}(W_{\mathcal{O}})$, and they are called the unobservability index and estimation condition number, respectively \cite{Krener2009}. 
If $\mathcal{O}_\varepsilon$ has full column rank, then the singular values of $\mathcal{O}_\varepsilon$ approach the square root of the eigenvalues of $\frac{1}{\Delta t}W_{\mathcal{O}}$ for small $\varepsilon,\Delta t$. Since the method presented in this paper analyzes $\mathcal{O}_\varepsilon$, we apply these established measures to the singular values of $\mathcal{O}_\varepsilon$ and focus on $\kappa(\mathcal{O}^{\top}_\varepsilon\mathcal{O}_\varepsilon)=\kappa(\mathcal{O}_\varepsilon)^2=[\bar{\sigma}(\mathcal{O}_\varepsilon)/\ubar{\sigma}(\mathcal{O}_\varepsilon)]^2$ for brevity. 

\section{Motivating examples}

To illustrate the challenges associated quantifying observability for individual state variables, consider the following series of examples. 

\begin{equation*}
\small
\begin{aligned}
\prescript{}{1}{\mathcal{O}}_\varepsilon = \begin{bmatrix}
        10^{-4} & 1 & 0 \\
        0 & 1 & 0 \\
    \end{bmatrix},
\quad  %\Sigma_\text{DT-NLTI}:\quad
\prescript{}{2}{\mathcal{O}}_\varepsilon = \begin{bmatrix}
        1 & 1 & 0 \\
        0 & 1 & 0 \\
    \end{bmatrix}.\\
\end{aligned}
\end{equation*}
In neither case is the system full rank. To assess the observability of the first state we could augment each system with the first state basis vector ($\vect{e}_1^\top = \begin{bmatrix}
    1&0&0
\end{bmatrix}$) and check to see if the rank has changed (as in \cite{VanBreugel2021a}). In both cases the rank does not change, suggesting that the first state is observable, however, it is clearly \textit{more} observable in $\prescript{}{2}{\mathcal{O}}_\varepsilon$, and a quantification of this difference would be helpful.

Established approaches for quantifying the level of observability involve looking at the eigenvalues of the Gramian. To see the challenges associated with these methods, consider the following. 
%hypothetical examples of $\mathcal{O}_\varepsilon$:
\begin{equation*}
\small
\begin{aligned}
\prescript{}{3}{\mathcal{O}}_\varepsilon = \begin{bmatrix}
        1 & 0 \\
        0 & 10 \\
    \end{bmatrix},
\quad  %\Sigma_\text{DT-NLTI}:\quad
\prescript{}{4}{\mathcal{O}}_\varepsilon = \begin{bmatrix}
        1 & 0 \\
        1 & 1 \\
    \end{bmatrix}, 
\quad  %\Sigma_\text{DT-NLTI}:\quad
\prescript{}{5}{\mathcal{O}}_\varepsilon = \begin{bmatrix}
        1 & 10^{-16} \\
        0 & 10^{-16} \\
    \end{bmatrix}.\\
% \begin{bmatrix}
%         0 & 1 \\
%         1 & 0 \\
%     \end{bmatrix}\begin{bmatrix}
%         100 \\
%         1 \\
%     \end{bmatrix},
% \quad  %\Sigma_\text{DT-NLTI}:\quad
% \begin{bmatrix}
%         -0.85 & -0.52 \\
%         -0.52 & 0.85 \\
%     \end{bmatrix}\begin{bmatrix}
%         2.6 \\
%         0.38 \\
%     \end{bmatrix}, 
% \quad  %\Sigma_\text{DT-NLTI}:\quad
% \begin{bmatrix}
%         -1 & -10^{-16} \\
%         -10^{-16} & 1 \\
%     \end{bmatrix}\begin{bmatrix}
%         1 \\
%         10^{-32} \\
%     \end{bmatrix}.
\end{aligned}
\end{equation*}
% \begin{equation}
% \tiny
% \mathcal{O}_1 = \begin{bmatrix}
%         1 & 1 & 0 \\
%         0 & 1 & 0 \\
%         0 & 0 & 10 \\
%     \end{bmatrix},
% \quad  %\Sigma_\text{DT-NLTI}:\quad
% \mathcal{O}_2 = \begin{bmatrix}
%         1 & 1 & 0 \\
%         0 & 1 & 0 \\
%         0 & 1 & 1 \\
%     \end{bmatrix}, 
% \quad  %\Sigma_\text{DT-NLTI}:\quad
% \mathcal{O}_3 = \begin{bmatrix}
%         1 & 1 & 10^{-16} \\
%         0 & 1 & 10^{-16} \\
%         0 & 0 & 10^{-16} \\
%     \end{bmatrix}.
% \end{equation}
Estimating the first state variable ($j=1$) is equally well-posed for $\prescript{}{3}{\mathcal{O}}_\varepsilon$ and $\prescript{}{4}{\mathcal{O}}_\varepsilon$, and practically, $\prescript{}{5}{\mathcal{O}}_\varepsilon$ too. However, the condition numbers of the Gramians are all different. Investigating the rows of the Gramians does not provide additional insight either, e.g. $\scriptsize \prescript{}{4}W_{\mathcal{O}_\varepsilon}=\begin{bmatrix}
    2 & 1 \\ 1 & 1\end{bmatrix}$ hides the fact that the first state is \textit{directly} observable. The second row of $\prescript{}{4}W_{\mathcal{O}_\varepsilon}$ does not actually provide any information about the first state (unless there is a different measurement, or there are dynamics, that make it possible to decouple this combined measurement into its components). 
%Since the eigenvectors of $\prescript{}{2}{W}_\mathcal{O}$ each mix both states $\{[\shortminus0.85, \shortminus0.52]^\top, [\shortminus0.52, 0.85]^{\top}\}$, 
%This makes it impossible to use the eigenvalues to determine an observability metric for just the first state in this orthogonal space. 
To summarize, each row of $\mathcal{O}_\varepsilon$ that does not contribute to closely reconstructing the basis vector $\vect{e}_j$ can confound efforts to quantify the observability of the $j^{th}$ state. Thus, we develop an approach for finding a small subset of $\mathcal{O}_\varepsilon$ that is sufficient for reconstructing $\vect{e}_j$, and which yields a small condition number. There are, however, likely cases where in practice (e.g. with noisy sensors) using more rows of $\mathcal{O}_\varepsilon$ (e.g. more sensors, or more measurements in time) will yield a better state estimate for the $j^{th}$ state, despite corresponding to a larger condition number compared the small subset of rows that we select. 

% \newpage
\section{Empirical Individual State Observability (E-ISO)}

E-ISO provides measures of (un)observability for individual state variables by quantifying how well-posed the problem of estimating a single state variable is given measurements and inputs from a time window of length $w$. This process involves finding the combination of sensors and measurements, i.e. rows of $\mathcal{O}_\varepsilon$, that provide the “best” value for the chosen measure (e.g. the smallest condition number). Solving this problem with a brute force approach would be computationally intractable for large systems, as this is equivalent to finding the best possible combination of rows in $\mathcal{O}_{\varepsilon}$ for observing the state variable of interest. The number of combinations that would need to be evaluated for such an approach is given by
\begin{equation} \label{comb}
\small
N =\sum_{r=1}^{pw}\binom{pw}{r} = \sum_{r=1}^{pw} \frac{(pw)!}{r!(pw - r)!},
\end{equation}
where $pw$ is the number of rows in $\mathcal{O}_{\varepsilon}$. For perspective, in a system with four outputs and 25 simulation steps ($pw=100$) there are $N>10^{30}$ combinations. 

This section details an efficient, but approximate, solution consisting of three steps. First, a sparse subset of rows is selected ($\mathcal{O}_\varepsilon^{\vect{e}_j}$) whose linear combination can reconstruct $\vect{e}_j$ to within a user-specified tolerance ($\beta$). Second, (un)observability measures of the corresponding approximate observable subspace ($\hat{\mathcal{O}}_\varepsilon^{\vect{e}_j}$) are evaluated.  
%We then calculate approximate (un)observability metrics for the state of interest by analyzing the singular values of $\hat{\mathcal{O}}_\varepsilon^{\vect{e}_j}$. 
Finally, since many unique subsets of rows of $\mathcal{O}_\varepsilon$ can be found to reconstruct $\vect{e}_j$, these unique subsets are sequentially gathered, in order of decreasing sparsity, into a collection $^i\mathcal{O}_\varepsilon^{\vect{e}_j}$ that increases in size with each iteration ($i$). For each iteration, (un)observability measures of the approximate observable subspace are calculated. The final measure that describes the approximate observability of the state variable of interest is the “best” of these. %, e.g. using the condition number as a metric the minimum condition number of $^i\hat{\mathcal{O}}_\varepsilon^{\vect{e}}$ is chosen for $i=\{0,1,2,...,w\}$. 
Pseudo-code is provided at the end of the section.

\subsection{Selecting a sparse subset of the observability matrix}
\begin{figure}[tbh]
\centerline{\includegraphics{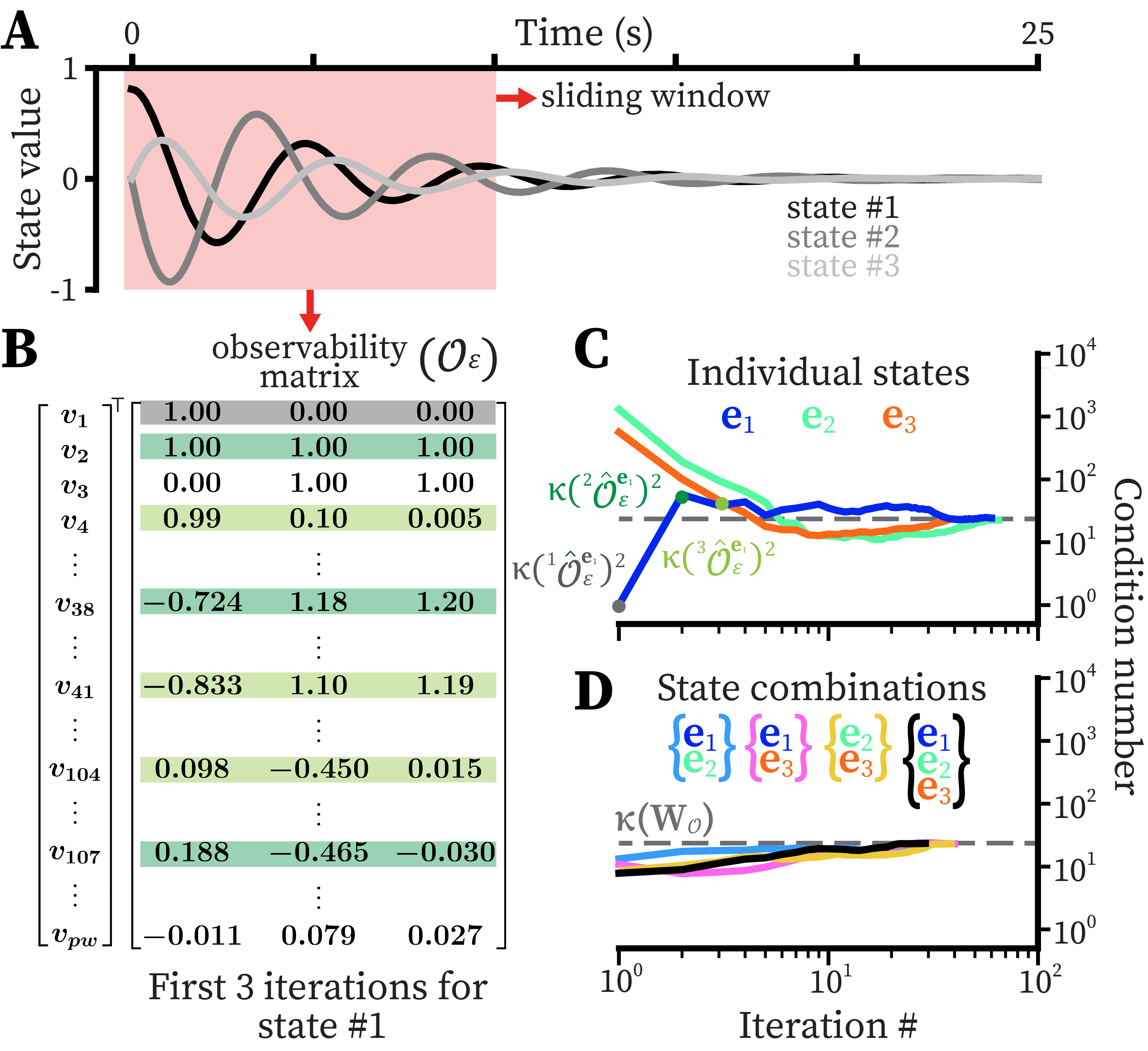}}
\caption{Graphical illustration of E-ISO applied to a fully observable linear system with the dynamics $\dot{\vect{x}}=A\vect{x} = \colvec[0.5]{ 0 & 1 & 0 \\ -2 & 0 & 1 \\ 1 & 0 & -1}\vect{x}$, $\vect{y}=\colvec[0.5]{ 1 & 0 & 0 \\ 1 & 1 & 1 \\ 0 & 0 & 1}\vect{x}$. \textbf{A.} Simulation from which the empirical observability matrix ($\mathcal{O}_\varepsilon$) is constructed over time in a sliding window ($w=100$). \textbf{B.} Example $\mathcal{O}_\varepsilon$ (Eq. \ref{DT-EOM}) and free parameter vector $\vect{v}$ (Eq. \ref{s_recon}) used for optimization. The rows of $\mathcal{O}_{\varepsilon}$  selected from the first three optimization iterations of E-ISO are highlighted. \textbf{C.} The condition number of the observable subspace of each iteration of E-ISO for each individual state variable, which converge to the condition number of the observability Gramian (gray dashed line). \textbf{D.} Same as C, but for combinations of state variables. E-ISO parameters: $\alpha=10^{-2}, \beta=10^{-3}, \sigma_0=10^{-6}$.}
\label{fig:O_method}
\end{figure}

For a state to be observable given $\mathcal{O}_\varepsilon$, it must be possible to linearly combine the rows to reconstruct the basis state vector corresponding to the state variable of interest ($\vect{e}_j$ for the $j^{th}$ state variable). That is, there must be a vector $\vect{v}$ such that
\begin{equation} \label{s_recon}
\small
\vect{e}_j^\top = \vect{v}^\top\mathcal{O}_{\varepsilon}. 
\end{equation}
%where
%\begin{equation} \label{v}
%\vect{v} = \begin{bmatrix} v_1 & v_2 & \hdots & v_{p \times w} \end{bmatrix}.
%\end{equation}
%If it is possible to choose $\vect{v}$ such that $\vect{e}_k$ is reconstructed, within some tolerance, then it can be said that the $j_{th}$ state variable is \textit{approximately} observable.
We define $\mathcal{O}_\varepsilon^{\vect{e}_j}$ as the subset of $\mathcal{O}_{\varepsilon}$ corresponding to non-zero elements in $\vect{v}$. To efficiently exclude as many rows from $\mathcal{O}_\varepsilon^{\vect{e}_j}$ as possible, we use established optimization tools \cite{cvxpy,mosek} to find $\vect{v}_o$ that minimizes a constrained convex problem,
\begin{equation} \label{eqn:loss}
\small
\vect{v}_o = 
\begin{aligned}
\arg\min_{\vect{v}} \quad & || \vect{e}_j - \mathcal{O}_{\varepsilon}^\top{\vect{v}} ||_2 +  \alpha || \vect{v} ||_1\\
\textrm{s.t.} \quad & |(\vect{e}_j - \mathcal{O}_{\varepsilon}^\top{\vect{v}}^{\top})_s| \leq \beta, ~ s=\{1,2,...,n\} \\
%\quad & \beta>0
%\vect{v}_o = \underset{\vect{v}}{\min} \enspace \mathcal{L}(\vect{v}, \vect{e}_j) \\
\end{aligned},
\end{equation}
where $\alpha$ is a scalar hyper-parameter, and $\beta>0$ is a tolerance on how closely each state element of $\vect{e}_j$ must be reconstructed. The objective function consists of two terms: 1) the $\ell_2$-norm of the reconstruction error, and 2) the $\ell_1$-norm of the free variable $\vect{v}$. The first term drives $\vect{v}$ to reconstruct the basis vector $\vect{e}_j$, whereas the second term is a regularizer that promotes sparsity in $\vect{v}$. Furthermore, the $\ell_1$-norm penalty on $\vect{v}$ also helps to prioritize the selection of rows of $\mathcal{O}_\varepsilon$ containing large values, thereby choosing rows that are likely to increase the magnitude of the singular values of $\mathcal{O}_\varepsilon$. In practice, the $\ell_1$-norm does not always drive small elements of $\vect{v}$ to zero. Thus, we add an extra step to eliminate small values in $\vect{v}_0$ by sequentially adding the largest elements until the tolerance set on $\vect{e}_j$ is met. Figure \ref{fig:O_method}C, gray shading, shows that this optimization selects a single row from $\mathcal{O}_\varepsilon$ to estimate $\vect{e}_1$. Rows highlighted in green and teal are discussed in Sec. \ref{sec:iterating}. If no solution to the optimization can be found, barring computational idiosyncrasies of the selected solver, we conclude that the system is \textit{approximately} unobservable.

%Thus, we add an extra step to sequentially eliminate small values in $\vect{v}_0$, starting with the smallest one, until this elimination causes the tolerance set on $\vect{e}_j$ to be exceeded. Figure \ref{fig:O_method}C, gray shading, shows that this optimization selects a single row from $\mathcal{O}_\varepsilon$ to estimate $\vect{e}_1$.%

\subsection{Obtaining a quantitative (un)observability measure}

%The optimization approach utilizes all of the rows of $\mathcal{O}_{\varepsilon}$ to determine a qualitative binary (observable or unobservable) measure of observability for a given state variable, or combination of state variables, but in many cases it is preferable to have a quantitative metric of observability. Prior approaches have employed quantitative metrics, such as the condition number of the observability Gramian \cite{Krener2009}, but these metrics have typically been limited to describe the observability of the entire system, rather than a individual state variable.

To compute quantitative (un)observability measures for individual state variables, the singular values of $\mathcal{O}_\varepsilon^{\vect{e}_j}$ can be analyzed. Since this subset may not have full column rank, an approximate observable column-subspace of $\mathcal{O}_\varepsilon^{\vect{e}_j}$ is found by calculating a rank-truncated singular value decomposition given a user-specified threshold ($\sigma_0$). We define the projection of $\mathcal{O}_\varepsilon^{\vect{e}_j}$ onto the approximately observable subspace as $\hat{\mathcal{O}}_\varepsilon^{\vect{e}_j}$. Now, established (un)observability measures, such as the condition number, can be applied to $\hat{\mathcal{O}}_\varepsilon^{\vect{e}_j}$ to obtain a quantitative (un)observability measure for the state variable of interest (Fig. \ref{fig:O_method}B--C, iteration \# = 1). 

\subsection{Quantifying observability for iterated subsets of $\mathcal{O}_\varepsilon$} \label{sec:iterating}

The optimization problem defined by Eq. \ref{eqn:loss} yields a single set of rows corresponding to an observable state variable, however, there may be multiple valid combinations of rows. %Only considering one subset may leave potentially important information out. 
To ensure that all relevant rows are accounted for, the optimization should ideally be iterated for every possible combination of rows in $\mathcal{O}_{\varepsilon}$, and then the subset of selected rows with the minimum condition number could  be used to evaluate the observability of the state variable of interest. The first optimization of Eq. \ref{eqn:loss} yields $^1\mathcal{O}_{\varepsilon} ^{\vect{e}_j}$ (Fig. \ref{fig:O_method}B, gray row). The rows in $^1\mathcal{O}_{\varepsilon} ^{\vect{e}_j}$ are then removed from $\mathcal{O}_{\varepsilon}$ and the optimization is repeated to select new rows (Fig. \ref{fig:O_method}B, green rows), which are added to $^1\mathcal{O}_{\varepsilon}^{\vect{e}_j}$ and the new collection is defined as $^2\mathcal{O}_{\varepsilon}^{\vect{e}_j}$. This process is repeated (e.g. Fig. \ref{fig:O_method}B, teal rows) until the optimization fails to find an observable subset. At each iteration the condition number of $^i\hat{\mathcal{O}}_{\varepsilon}^{\vect{e}_j}$ is determined (Fig. \ref{fig:O_method}C). To obtain the best single measure, the minimum condition number is selected:
\begin{equation} \label{O_iter}
\small
{\kappa}_{\min} = \min \{ \kappa(^1\hat{\mathcal{O}}_{\varepsilon}^{\vect{e}_j})^2, \hspace{2pt} \kappa(^2\hat{\mathcal{O}}_{\varepsilon}^{\vect{e}_j})^2, \hspace{2pt} \hdots \hspace{2pt}, \hspace{2pt} \kappa(^w\hat{\mathcal{O}}_{\varepsilon}^{\vect{e}_j})^2 \},
\end{equation}
where $\kappa(^i\hat{\mathcal{O}}_{\varepsilon}^{\vect{e}_j})$ is the condition number of the collection of rows selected from $\mathcal{O}_{\varepsilon}$ after $i$ iterations. For directly measurable states, the smallest $\kappa$ will occur at the first iteration (Fig. \ref{fig:O_method}C, $\vect{e}_1$); for states requiring an accumulation of sensor measurements (Fig. \ref{fig:O_method}C, $\vect{e}_{2}$ and $\vect{e}_{3}$) the smallest $\kappa$ typically occurs at some intermediate iteration number.

% To ensure that the computed condition numbers are consistent with those from the observability Gramian, the square of $\kappa(^i\hat{\mathcal{O}}_{\varepsilon}^{\vect{e}_j})$ is used, which is equivalent to $\kappa((^i\hat{\mathcal{O}}_{\varepsilon}^{\vect{e}_j})^\top (^i\hat{\mathcal{O}}_{\varepsilon}^{\vect{e}_j}))$. 

%\subsection{Convergence to empirical observability Gramian metrics}
The E-ISO method can be extended to determine a single observability measure for any combination of $z$ states by stacking $z$ unique $\vect{e}_j$'s and defining $\vect{v}$ as a $pw \times z$ matrix (Fig. \ref{fig:O_method}D). For a fully observable linear system, the condition number $\kappa(^i\hat{\mathcal{O}}_{\varepsilon}^{\vect{e}_j})^2$ for each individual state, or any combination of states, will converge to the condition number of the observability Gramian $\kappa(W_\mathcal{O})$ after many iterations (Fig. \ref{fig:O_method}C--D), provided that every row of $\mathcal{O}_\varepsilon$ is eventually selected. For some systems, our iterative algorithm will not, however, eventually select every row. 

Pseudo-code for E-ISO, which we implemented in Python, is provided below.

\begin{algorithm}
\small
\hspace*{\algorithmicindent} \textbf{Input:} $\mathcal{O}_{\varepsilon}$, $\vect{e}_j$ 
\hspace*{\algorithmicindent} \textbf{Parameters: $\alpha$, $\beta$, $\sigma_0$} 
\hspace*{\algorithmicindent} \textbf{Output: $\kappa_{\min}$}
\caption{E-ISO}
\begin{algorithmic}[1] % [1] provides numbers in every line

    \State $\hat{\mathcal{O}}_\varepsilon^{\vect{e}_j} \gets (), \kappa \gets (), i \gets 1$ \Comment{initialize variables}
    % \State $\kappa \gets ()$  \Comment{empty, to store condition \#}
    % \State $i \gets 1$ \Comment{iteration counter}
    \State $IsObservable \gets$ True
    \While {$IsObservable$}
        \Statex \LeftComment{1} { reconstruct state with convex optimization}
        \Statex \LeftComment{1} {  $r$ is the collection of row indices used}
        \State $IsObservable, r \gets \Call{Optimize}{\mathcal{O}_\varepsilon, \vect{e}_j, \alpha, \beta}$
    \If {$IsObservable$}
        \State $\hat{\mathcal{O}}_\varepsilon^{\vect{e}_j} \gets [\hat{\mathcal{O}}_\varepsilon^{\vect{e}_j} , \mathcal{O}_\varepsilon(r, :)]$ \Comment{add rows to subset}
        \State $\kappa(i) \gets \Call{ConditionNumber}{\hat{\mathcal{O}}_\varepsilon^{\vect{e}_j}, \sigma_0}$
    \Else
        \If {$i = 1$} \Comment{failed on 1st iteration}
            \State $\kappa_{\min} \gets \infty$  \Comment{condition \# is undefined}
        \Else
            \State $\kappa_{\min} \gets \min (\kappa)$  \Comment{minimum condition \#}
        \EndIf
    \EndIf
        \State ${\mathcal{O}}_\varepsilon(r, :) \gets 0$ \Comment{set rows to zero}
    \State $i \gets i + 1$ \Comment{next iteration}
    \EndWhile
      
\end{algorithmic}
\end{algorithm}

% \clearpage
\section{Applications}

We apply the E-ISO approach to two examples: a linear system to highlight sensor selection applications, and a nonlinear biological system to highlight applications to trajectory planning for active sensing.

\subsection{Sensor selection}
\begin{figure}[hbt!]
\centerline{\includegraphics{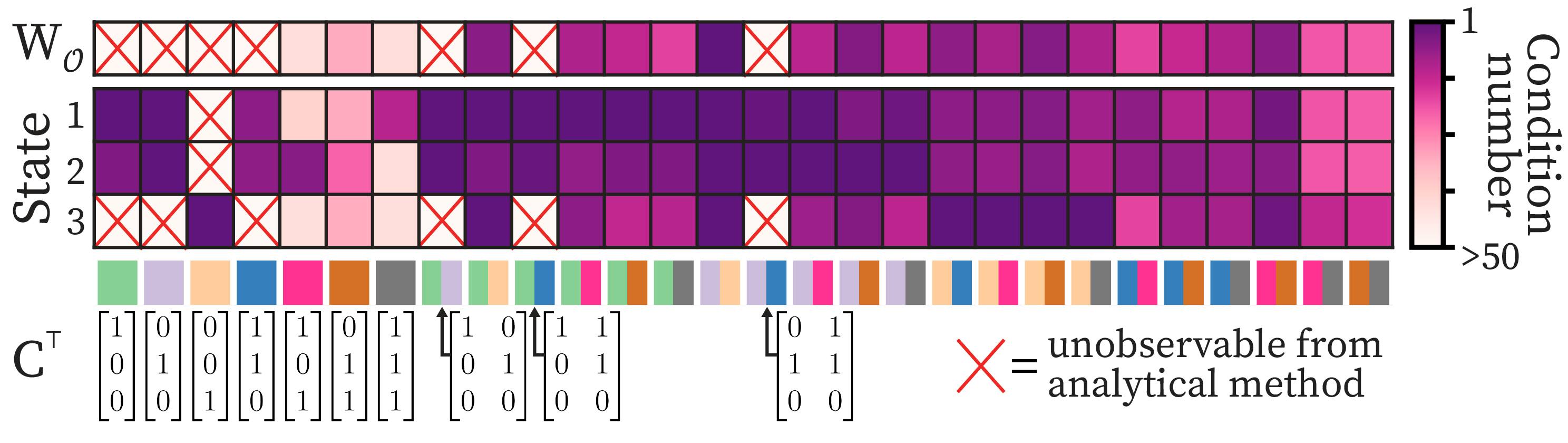}}
\caption{The condition number of the observability Gramian (top) is generalized to individual state variables with E-ISO (middle). Different sensor sets are represented by $C$ matrices (bottom) applied to a discrete-time linear system with dynamics ${\vect{x}_{k+1}}=A\vect{x}_k$, $\vect{y}_k=C\vect{x}_k$ where $A = \colvec[0.5]{ 0.9952 & 0.095 & 0 \\ -0.095 & 0.9002 & 0 \\ 0 & 0 & 0.9048 }$. E-ISO parameters: $\alpha=10^{-4}, \beta=10^{-2}, \sigma_0=10^{-6}$ multiplied by maximum eigenvalue.}
\label{fig:sensor_selection}
\end{figure}

%For dynamical systems, it is advantageous to directly measure state variables to simplify state estimation. However, this is not always practical for high-dimensional systems, which has motivated development of sparse sensing approaches \cite{manohar2018data}. Applications concerned with full sate estimation have focused on metrics related to the observability Gramian, such as maximizing the rank \cite{Bartos2021}. But, sparse sensing approaches to optimize observability of individual state variables are lacking.
The need for efficiently estimating states of high dimensional systems while minimizing the quantity of physical sensors has spurred the development of sparse sensing approaches \cite{manohar2018data, Brace2022}, but these methods focus on full state estimation. Here, E-ISO is applied to a simple discrete-time linear system with various output configurations to illustrate how sensors could be chosen to maximize the observability of an individual state variable of interest. The analysis shows that the condition number of the Gramian is correlated with the least observable state variable ($R^2 = 0.99$)---thus obscuring information about the more observable state variables---whereas E-ISO can resolve differences in observability between state variables (Fig. \ref{fig:sensor_selection}).

\subsection{State trajectory planning for active sensing}

For nonlinear systems, the observability of the state variables can depend on the current state, and some non-zero inputs may be required to guarantee observability \cite{Kunapareddy2018}. To illustrate E-ISO's application to such \textit{active sensing} objectives, consider the following system inspired by a flying insect such as a fruit fly \cite{VanBreugel2021a}:
\begin{equation} \label{fly_wind_model}
\small
\vect{\dot{x}} = \begin{bmatrix} \dot{d} \\ \dot{g} \\ \dot{w} \\ \dot{\phi} \\ \dot{\zeta} \\ \end{bmatrix}
=
\begin{bmatrix} 0 \\ u_g \\ 0 \\ u_{\phi} \\ 0 \\ \end{bmatrix}
, \quad
\vect{h}(\vect{x}(t)) = \begin{bmatrix} \phi \\ g/d  \\ \gamma \end{bmatrix}.
\end{equation}

States 1--3 represent magnitudes describing the fly's altitude $d$, ground speed $g$, and the ambient wind speed $w$. States 4--5 are angular quantities describing the fly's heading $\phi$ and the ambient wind direction $\zeta$. For simplicity heading and course direction are defined to be equal. In this example, $d$, $w$, and $\zeta$ are constant, whereas $\phi$ and $g$ are directly controlled with inputs $u_{\phi}$ and $u_g$. All dynamics (inertial, aerodynamics, etc.) are excluded to simplify the presentation of the observability analysis. The nonlinear outputs $\vect{h}(\vect{x}(t))$ of the system consist of $\phi$ measured directly, optic flow approximated by $g/d$ \cite{VanBreugel2014a}, and the air speed angle in the global frame:
\begin{equation} \label{air_speed}
\small
\gamma = \arctan \left( \frac{-g \sin{\phi} + w \sin{\zeta}}{-g \cos{\phi} + w \cos{\zeta}} \right).
\end{equation}
For a fly engaged in a chemical plume tracking behavior, estimating $\zeta$ is especially important \cite{van2014plume}. Thus, $\zeta$ can be considered a high-priority estimate, whereas $g, d, w$ are not needed for most plume tracking algorithms. 
%Due to the nonlinear outputs, $\zeta$ and the other states may at some times be unobservable; E-ISO can reveal which inputs must be applied to recover observability of $\zeta$. 
E-ISO can reveal which inputs are needed to ensure observability of an individual state variable, such as $\zeta$.  
The following E-ISO results confirm prior analytical work and its extension to the exact dynamics given above (\cite{VanBreugel2021a}, see also: \textit{https://github.com/BenCellini/EISO}).

\begin{figure*}[tb]
\centerline{\includegraphics{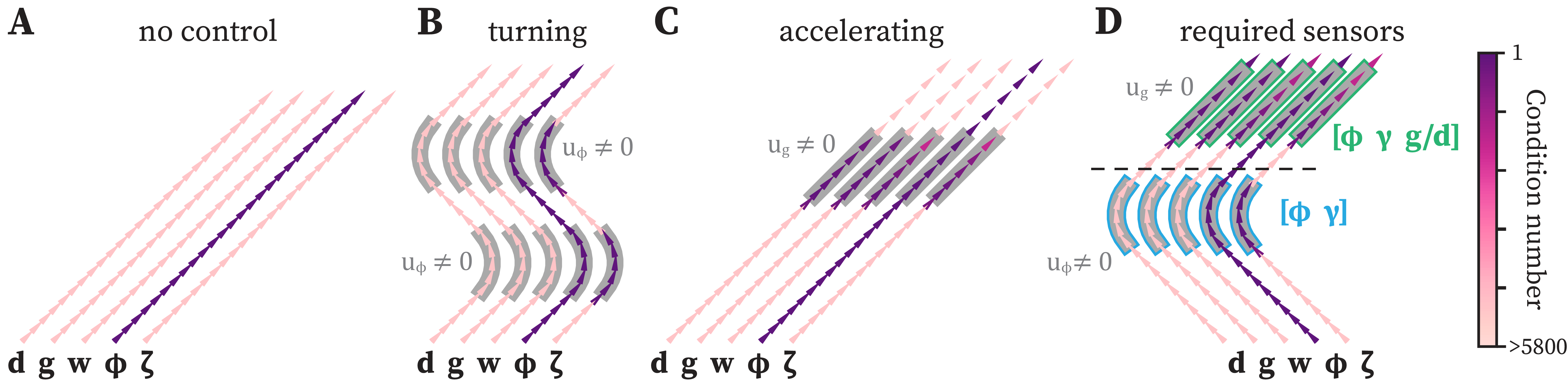}}
\caption{E-ISO reveals flies must turn or accelerate to estimate wind direction. \textbf{A.} Simulated five-second trajectory from Eq. \ref{fly_wind_model} with constant optic flow, no turning, and $\Delta t = 0.1$ . For each individual state variable in Eq. \ref{fly_wind_model}, color shading indicates the observability level (Eq. \ref{O_iter}) for sliding windows ($w=3$). E-ISO parameters: $\alpha=10^{-6}, \beta=10^{-3}, \sigma_0=10^{-8}$. \textbf{B.} Same as A, but for a trajectory with turns. \textbf{C.} Same as A, for a trajectory with translational acceleration. %Gray shading indicates control inputs were on. 
\textbf{D.} Different sensors from Eq. \ref{fly_wind_model} are needed to estimate $\zeta$ for different trajectories: only $\phi$ and $\gamma$ are necessary when turning, but $g/d$ is also required when accelerating.}
\label{fig:state_trajectory}
\end{figure*}

First, the observability of all the state variables was evaluated for a non-zero constant optic flow trajectory $\dot{g} = 0$ with zero inputs $u_g = u_{\phi} = 0$. Only $\phi$, a direct measurement, is observable (Fig. \ref{fig:state_trajectory}A). %This suggests that an active sensing strategy may be necessary to estimate $\zeta$ and/or the other states.

Prior work has shown that flies perform rapid turns called saccades at a rate of $\SI{0.5}{\hertz}$ during flight \cite{Cellini2020}, which could potentially improve their ability to estimate the ambient wind direction \cite{VanBreugel2021a, VanBreugel2022}. In Eq. \ref{fly_wind_model}, saccades can be emulated by introducing a nonzero change in heading $\dot{\phi} \neq 0$ by setting $u_{\phi} \neq 0$. Applying E-ISO to the resulting trajectories reveals that $\zeta$ becomes observable, but the rest of the states (aside from $\phi$) remain unobservable (Fig. \ref{fig:state_trajectory}B).

%Turning trajectories recover the observability of $\zeta$ (Fig. \ref{fig:state_trajectory}B), but what about the other unobservable sates? 
Prior work has also shown that with monocular optic flow measurements $g/d$, both $d$ and $g$ are only observable when a known translational acceleration $\dot{g} \neq 0$ is applied (i.e. $u_{g} \neq 0$ in Eq. \ref{fly_wind_model}) \cite{ VanBreugel2014a}, \mbox{\cite{Lingenfelter2021}}. Applying E-ISO to accelerating trajectories revealed that all the state variables become observable (Fig. \ref{fig:state_trajectory}C). Together, the E-ISO results suggest that flies could use two distinct active sensing strategies, turning and/or accelerating, to estimate $\zeta$---but flies would require acceleration to observe the full state (this result does require any model parameters to be calibrated \cite{VanBreugel2021a}). 

E-ISO can also identify which sensor combinations are necessary to observe an individual state variable given some trajectory. 
%This was applied to compare the full sensor set Eq. \ref{fly_wind_model} to a sensor set with just angular measurements $\vect{h}(\vect{x}(t))  = \begin{bmatrix} \phi & \gamma \end{bmatrix}$. 
 E-ISO shows that the angular sensor set alone is enough to estimate $\zeta$ for turning trajectories, but the full sensor set is still required to estimate the full state when accelerating (Fig. \ref{fig:state_trajectory}D).

\section{Discussion}

% Variations:
% \begin{itemize}
%   \item Could obtain Oe from analytical expressions
%   \item constructibility — burak’s textbook? this paper? https://ieeexplore.ieee.org/abstract/document/8827581, also possibly relevant: 
%     \item alternative metrics, and the value of the observability matrix for evaluating other metrics: how many sensors are required, data over what period of time, how many rows = how much memory
% \end{itemize}
\textit{Variations.} Although the narrative and examples presented here focus on the discrete-time empirical observability matrix as the starting point, trivial modifications include starting from an analytically determined observability matrix, %$\prescript{w}{}{\mathcal{O}=d\mathcal{G}}$, where $d\mathcal{G}$ is given by Eqns. \ref{ctMapping},\ref{dtMapping}
or a constructability matrix. Furthermore, although we focus on the condition number, alternative measures such as the unobservability index can be used instead. In fact, by working with $\mathcal{O}_\varepsilon$, instead of the Gramian $W_\mathcal{O}$, additional measures can be explored including: how many distinct sensors are necessary, how many measurements are needed, and the size of the time window required to capture the level of observability along state trajectories.
%E-ISO can also be extended to evaluate the observability of nonlinear combinations of state variables by setting the matrix to reconstruct $\vect{e_j}$ as the Jacobian of the nonlinear function evaluated at the initial state. As opposed to analytical methods, which would require transforming the state variables into a new coordinate space to evaluate the observability of non-state variables, this empirical approach can evaluate the observability of nonlinear combinations  of state variables with the original dynamical model. Many of the core elements of E-ISO also apply to evaluating controllability, in which a similar optimization routine to E-ISO can be applied to analyze the controllability of individual state variables from an analytically or empirically constructed controllability matrix.

% \begin{itemize}
%   \item make a note of scaling recommendations
%   \item computational expense for high dimensional systems
%   \item hyper parameter tuning
%   \item connect to most observable directions in the gramian: fundamental problem: the SVD only produces orthonormal vectors, that imposes a constraint that we don't care about. Could relate to QR here?
% \end{itemize}
\textit{Limitations and practical recommendations.} We recommend scaling state variable units such that all states are as comparable in magnitude over time as possible, and scaling outputs according to their expected noise levels. Normalizing the outputs is not recommended, as this would change the interpretation of the singular values of $\hat{\mathcal{O}}_\varepsilon^{\vect{e}_j}$ with respect to the states themselves. E-ISO has three hyper-parameters that may require tuning, or a methodical sweep. When state values are scaled from $0.1$ to $10$, we recommend starting with: $\alpha=10^{-6}, \beta=10^{-3}$. For systems with a large number of states and/or measurements, we suggest increasing $\alpha$ and $\beta$ to find sparse sets of $\mathcal{O}_\varepsilon$, at the expense of reconstruction tolerance. When choosing the singular value threshold ${\sigma}_{0}$, for which to calculate the rank-truncated condition number, it is generally critical to pick a value that excludes any singular values that are not required to reconstruct the state variable of interest. As a test, a rank-truncated observability matrix can be constructed from a subset of the singular values of the full observability matrix $\mathcal{O}_\varepsilon$. This procedure can be iterated starting from the largest singular value until the state variable of interest can be reconstructed within the $\beta$ tolerance from the rank-truncated observability matrix---and the last value of the smallest singular value required can be set as the ${\sigma}_{0}$ threshold. Without this routine, an arbitrary chosen ${\sigma}_{0}$ can lead to inaccurate condition numbers. When possible, for smaller systems, this method of choosing ${\sigma}_{0}$ combined with evaluating every possible combination of rows will yield the most accurate condition numbers for individual states variables. We provide an example of this implementation on the E-ISO GitHub. E-ISO can become computationally cumbersome for large systems. Rather than implement E-ISO in real-time for trajectory planning, we recommend using E-ISO to identify trajectory motifs a priori (e.g. turning or accelerating), and using these motifs to rank active sensing trajectories.

%\newpage
% Applications:
% \begin{itemize}
%   \item relevance to machine learning literature: data selection for training
%   \item partial update kalman filters https://ieeexplore.ieee.org/abstract/document/9626946
%   https://ieeexplore.ieee.org/abstract/document/9561008
%   \item tool for understanding active sensing in animals: cite some high profile active sensing related stuff across organisms?
%\end{itemize}  

\textit{Applications.} E-ISO provides a practical solution to the open problem of methodically discovering trajectories that guarantee observability of specific state variables (or parameters) in partially observable (nonlinear) systems \cite{mania2022active}. E-ISO's focus on individual state variables can serve as a practical generalization to full state methods such as empirical Gramian-based observability methods, and sparse sensor selection algorithms \cite{manohar2018data}. In particular, E-ISO goes beyond established methods like Kalman canonical decomposition, which can be used to find the observable subspace, but does not single out individual state variables within that subspace. Beyond sensor selection and trajectory planning, E-ISO can also be used to curate data prior to building bespoke observers using data-hungry machine learning methods to limit extraneous/unobservable information. E-ISO results could also be incorporated into partial update Kalman filters that use observability measures to throttle state estimation \cite{humberto2021}. Finally, E-ISO can be used as an elegant analysis and hypothesis generation tool for understanding active sensing in biological systems that may not be concerned with full state estimation.

%These findings demonstrate a useful application of resolving the differences in observability between individual state variables, as opposed to only considering the observability of the full systems (which in this case is unobservable). Ideally, E-ISO can be applied to provide insights that might not otherwise be apparent from traditional observability tools.

%%%%%%%%%%%%%%%%%%%%%%%%%%%%%%%%%%%%%%%%%%%%%%%%%%%%%%%%%%%%%%%%%%%%%%%%%%%%%%%%

%%%%%%%%%%%%%%%%%%%%%%%%%%%%%%%%%%%%%%%%%%%%%%%%%%%%%%%%%%%%%%%%%%%%%%%%%%%%%%%%

%%%%%%%%%%%%%%%%%%%%%%%%%%%%%%%%%%%%%%%%%%%%%%%%%%%%%%%%%%%%%%%%%%%%%%%%%%%%%%%%
%\section*{APPENDIX}

\section*{Funding Sources}
This work was partially supported by funding from the Air Force Office of Scientific Research (FA9550-21-0122), the Sloan Foundation (FG-2020-13422), and the National Science Foundation AI Institute in Dynamic Systems (2112085). 

\section*{Acknowledgement}

The authors thank Richard Murray, Trevor Avant, Natalie Brace, Aditya Nair, Petros Voulgaris, and Noah Cowan for feedback on the manuscript.

\bibliographystyle{plainnat}
\bibliography{references}  %%% Uncomment this line and comment out the ``thebibliography'' section below to use the external .bib file (using bibtex) .

\end{document}